\documentclass[twocolumn,preprintnumbers,amsmath,amssymb,groupedaddress,prl]{revtex4}

\usepackage{graphicx}
\usepackage{dcolumn}
\usepackage{bm}
\usepackage{epsfig}

\def\e{\begin{equation}}
\def\f{\end{equation}}
\def\=#1{\overline{\overline #1}}

\def\-#1{{\bf #1}}
\def\.{\cdot}

\begin{document}

\title{Comment on ``Guiding, focusing, and sensing on the subwavelength scale using metallic wire arrays''}

\author{Pavel A. Belov$^{1}$, Atiqur Rahman$^{1}$, M\'{a}rio G. Silveirinha$^2$, Constantin R. Simovski$^{3}$, Yang Hao$^{1}$, Clive Parini$^{1}$}
\affiliation{$^1$Dept. Electronic Engineering, Queen Mary University
of London, Mile End Road, E1 4NS, London, UK\\
$^2$Universidade de Coimbra, Instituto de Telecomunica\c{c}\~{o}es,
3030 Coimbra, Portugal \\ $^3$Radio Lab., Heslinki University of
Technology, P.O. Box 3000, FIN-02015 HUT, Finland }
\begin{abstract}
\end{abstract}
\maketitle

In a recent Letter, Shvets et al \cite{ShvetsPendry} describe a
multiwire endoscope and claim that it is capable of guiding
electromagnetic field distributions preserving their subwavelength
details. The letter contains results of simulations for endoscope
consisting of $3\times 3$ array of wires and a claim that `a
practical multichannel endoscope will have a much larger (e.g. $25
\times 25$) number of metal wires'. We performed numerical
simulation of the $25 \times 25$ multiwire endoscope (see Fig.1a)
with exactly the same parameters as suggested by Shvets et al using
CST Microwave Studio package and didn't observe a satisfactory
imaging performance (see Fig.1d).

\begin{figure}[b] \vspace{-3mm}
\centering \epsfig{file=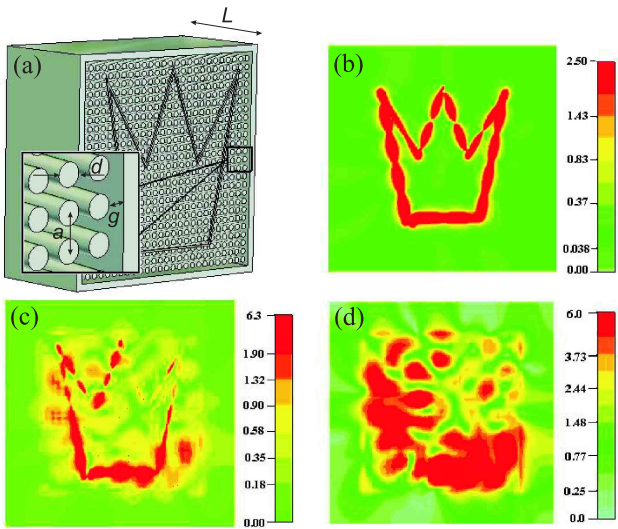, width=8.5cm} \caption{a)
Geometry of $25 \times 25$ multiwire endoscope proposed in
\cite{ShvetsPendry}. The wires have length $L=4\pi/3$ and diameter
$d=\lambda/15$. The period of array is $a=\lambda/10$. The shield is
placed at distance $g=\lambda/20$ from the marginal wires. The
nearfield source in the form of crown is placed at $0.07\lambda$
distance away from the endoscope. b) The amplitude of near field
created by the source at the input interface of the endoscope if the
interaction between the endoscope and the source is neglected. c)
The same distribution but in the presence of the endoscope. A huge
reflection and excitation of surface waves is observed. d) The
distribution of near field at the output interface of the endoscope.
No crown-like distribution as (b) is observed.} \label{Atiqur}
\end{figure}

Previous theoretical and experimental studies of multiwire
transmission devices \cite{canal,SWIWM, resolWM, APLWM, WMIR} which
were ignored in \cite{ShvetsPendry} allows us to conclude that
endoscope does not operate properly because of two reasons. Firstly,
the endoscope is coated by metal, when such kind of shielding
actually greatly reduces and limits its performance. In the absence
of shield the imaging performance of the endoscope would be greatly
increased. Secondly, an array of metallic wires can be efficiently
used for sensing, guiding and focusing in the subwavelength scale,
only if the length of the wires is tuned to obey the Fabry-Perot
(FP) resonance condition. This requirement is not fulfilled in
\cite{ShvetsPendry} whereas it is of fundamental importance in order
that the fields can be effectively sensed by the endoscope.

In subwavelength structures the electric field can be described to a
good approximation by an electric potential. The multiwire system
basically behaves as a ``sampler" of electric potential. Such
property implies that the period of the array roughly determines the
resolution of the system. However, the presence of the metal shield
forces the electric potential to be constant around a circumference
very close to the object to be imaged. This implies that the
electric field distribution near the considered object is corrupted
by the presence of the endoscope (see Fig. 1c). In the other words,
diffraction on the metal shield is extremely harmful to the image
formation. This problems can be avoided by using a multiwire
endoscope with no metallic shield and thickness tuned to obey FP
resonance condition. In this case, as demonstrated in \cite{YanOE},
because of the all-spatial-spectrum FP resonance the multiwire
endoscope does not produce any diffraction effects.

Basically, the multiwire system must be operated in the
``canalization'' regime described in \cite{canal}. The key idea is
to transform the whole spectrum of spatial harmonics generated by
the source, including evanescent waves, into propagating eigenmodes
of a metamaterial. This enables transmission of any field with
subwavelength details from the front interface of the metamaterial
slab to the back one provided the thickness of the slab obey the FP
resonance condition. Such regime was studied in detail at the
microwave \cite{canal, Pekkaexp, SWIWM, resolWM, APLWM}, infrared
\cite{Silv_Nonlocalrods,WMIR} and even visible \cite{layeredPRB}
domains. It has been used for realization of subwavelength imaging
by photonic crystals, and in this context it is also known as
self-collimation \cite{Collim,Li,GarciaPomar}, directed diffraction
\cite{Chien} and tunneling \cite{Kuo}. It is well established
\cite{SWIWM, resolWM, APLWM, WMIR, Silv_Nonlocalrods} that an array
of metallic wires can be operated in the canalization regime up to
infrared frequencies due to the extraordinary waveguiding properties
of the quasi-TEM mode supported by the metallic wires, and that such
effect is weakly dependent on losses and on the plasmonic properties
of the metal \cite{WMIR, Silv_Nonlocalrods}.

The FP condition enhances the sensing properties of the wires,
enables a nearly perfect transmission of the image, and guarantees
the absence of strong reflections from the endoscope, which
otherwise perturb the near field distribution of the source (see
Fig.1c). Numerous numerical and experimental studies \cite{resolWM,
APLWM, WMIR} show that if the frequency of operation is
significantly away from FP resonance then the image is severely
distorted by surface waves (see Fig.1d). This is a very general
behavior, a characteristic of arrays of parallel wires, ad of
tapered arrays, as shown in \cite{PekkaAPL} where a three fold
magnification was demonstrated using an array of $21\times 21$
wires.

\bibliography{shvets}
\end{document}